# High-pressure phase and transition phenomena in ammonia borane $NH_3BH_3$ from X-ray diffraction, Landau theory, and *ab initio* calculations

Yaroslav Filinchuk,[1]* Andriy H. Nevidomskyy,[2] Dmitry Chernyshov,[1] and Vladimir Dmitriev[1]

[1] *Swiss-Norwegian Beam Lines, European Synchrotron Radiation Facility, 6 rue Jules Horowitz, BP-220, 38043 Grenoble, France,*

[2] *Department of Physics and Astronomy, Rutgers University, 136 Frelinghuysen Road, Piscataway, NJ 08854, USA*

## Abstract

Structural evolution of a prospective hydrogen storage material, ammonia borane $NH_3BH_3$, has been studied at high pressures up to 12 GPa and at low temperatures by synchrotron powder diffraction. At 293 K and above 1.1 GPa a disordered *I4mm* structure reversibly transforms into a new ordered phase. Its *Cmc*$2_1$ structure was solved from the diffraction data, the positions of N and B atoms and the orientation of $NH_3$ and $BH_3$ groups were finally assigned with the help of density functional theory calculations. Group-theoretical analysis identifies a single two-component order parameter, combining ordering and atomic displacement mechanisms, which link an orientationally disordered parent phase *I4mm* with ordered distorted *Cmc*$2_1$, *Pmn*$2_1$ and *P*$2_1$ structures. We propose a generic phase diagram for $NH_3BH_3$, mapping three experimentally found and one predicted (*P*$2_1$) phases as a function of temperature and pressure, along with the evolution of the corresponding structural distortions. Ammonia borane belongs to the class of improper ferroelastics and we show that both temperature- and pressure-induced phase transitions can be driven to be of the second order. The role of N-H…H-B dihydrogen bonds and other intermolecular interactions in the stability of $NH_3BH_3$ polymorphs is examined.

## I. INTRODUCTION

Current interest in light hydrides is growing due to their potential use as hydrogen storage materials. Despite the high hydrogen content most of these compounds are very stable. This intensifies experimental efforts to destabilize light hydrides by chemical doping or catalysis, usually on the basis of "empirical rules" or by "trial and error" method. A necessary step towards understanding the stability and therefore towards a rational design of light hydrides is a study of a structural response due to changing temperature and pressure, which for the most of light hydrides has not yet been accomplished. Here we report such a study for ammonia borane $NH_3BH_3$, based on the X-ray powder diffraction data.

Experimental mapping of polymorphism for light hydrides by X-rays is not a trivial task due to the presence of weakly scattering hydrogen atoms; neutron elastic scattering on the other hand suffers from absorption on boron nuclei and incoherent scattering from hydrogen. As a matter of fact, theoretical modeling of light hydrides is not a trivial task either, as one can see from the apparent disagreement between the observed and theoretically predicted structures for $LiBH_4$ [1, 2] and $Mg(BH_4)_2$ [3, 4, 5].

In order to overcome these difficulties we present a study combining x-ray powder diffraction aiming to uncover phase transition phenomena and basic structural properties, the density functional theory (DFT) calculations to assist in determining a crystal structure, and a symmetry-based analysis to identify possible polymorphs and reconstruct a pressure-temperature phase diagram. A powder diffraction experiment, augmented by theoretical analysis, provides a detailed picture of the structural response to changing temperature and pressure. Practical application of this scheme to a hydrogen storage material, ammonia borane $NH_3BH_3$, is one of the goals of the present paper. It is a light-weight molecular crystalline compound with high hydrogen content and relatively low hydrogen desorption temperature, and it is considered to be a promising material for hydrogen storage [6, 7, 8]. A transition between a disordered tetragonal phase (*I4mm* [9, 10]), stable at ambient conditions, and an ordered low temperature orthorhombic structure (*Pmn*$2_1$ [10, 11]) has been reported earlier [9, 12, 13], however the evolution of the structure with temperature and pressure has not yet been studied. To the best of our knowledge, the experimental information on phase transitions in ammonia borane under pressure is limited to Raman spectroscopic studies and is contradictory: one phase transition was revealed at 0.8 GPa in Ref. [14], while a similar experiment reported in [15] indicates two transformations at ~0.5 and 1.4 GPa. The recent study [16] suggests three transitions, at 2, 5 and 12 GPa. The structure of high pressure polymorphs of $NH_3BH_3$ has not yet been determined experimentally, and the authors are not aware of any theoretical predictions of these phases. Filling in this gap is the other goal of the present work.

Ammonia borane also represents an interesting case of a dihydrogen bonding, N-H$^{\delta+}$...$^{\delta-}$H-B, largely defining the structure and dynamics of the compound at ambient pressure [17]. The dihydrogen bonds, i.e. protonic-hydridic H$^{\delta+}$...$^{\delta-}$H interactions, are known to have strength and directionality comparable with those found in conventional hydrogen-bonded systems [18, 19]. In this work we address their role in structure and stability of ammonia borane under high pressure.

The paper is organized as follows. First, we give all necessary information on experimental and theoretical techniques we use. Second, we present our findings of the new high-pressure phase of ammonia borane and characterize it by combining structure solution from powder diffraction data with DFT calculations. Then we show the thermal expansion and compressibility data obtained from the diffraction experiment as a function of temperature and pressure. Finally, all the structural information has been parameterized in the framework of unified phenomenological theory that provides a generic phase diagram together with evolution of corresponding order parameters.

## II. EXPERIMENTAL AND THEORETICAL METHODS

Commercially available NH$_3$BH$_3$ (>97% purity, Sigma-Aldrich) was checked for purity by synchrotron powder diffraction and proved to be a single-phase sample. All diffraction experiments were done at the Swiss-Norwegian Beam Lines of the ESRF. Diffraction patterns were collected using MAR345 image plate detector, and a monochromatic beam with the wavelength of 0.71171 or 0.69408 Å. The sample-to-detector distance (250 mm) and parameters of the detector were calibrated using NIST standard LaB$_6$. Two-dimensional diffraction images were integrated using Fit2D software [20].

<u>High-pressure diffraction experiments.</u> Finely ground samples were loaded into a diamond anvil cell (DAC) with flat culets of diameter 600 μm. The samples were loaded into a hole of 250 μm in diameter drilled in stainless steel gaskets pre-indented to 60-80 μm thickness; the beam was slit-collimated to 140×140 μm$^2$. Ruby provided a pressure calibration with precision of 0.1 GPa. No pressure-transmitting medium was used since pure ammonia borane provided good quasi-hydrostatic conditions, evidenced by a small broadening of the ruby fluorescence peaks. Diffraction measurements were performed up to a maximum pressure of

12.1 GPa, followed by a step-wise decompression down to the ambient pressure. At each pressure point a diffraction pattern has been measured over 180 seconds of the exposure time.

<u>Variable-temperature powder diffraction.</u> Fine powder of $NH_3BH_3$ was filled into a glass capillary of 0.5 mm diameter. Capillary was cooled from 300 K to 110 K at a 30 K per hour rate, while synchrotron powder diffraction data were collected *in-situ*. Temperature was controlled with an Oxford Cryostream 700+. Data collection time was 30 sec per image, followed by a readout during 83 sec. During each exposure the capillary was rotated by 30° in the same angular interval. 184 images were collected in total. Uncertainties of the integrated intensities σ(I) were calculated at each 2θ-point applying Poisson statistics to the intensity data, considering the geometry of the detector.

<u>DFT Calculations</u>. We used an *ab initio* plane–wave pseudopotential DFT method [21, 22] as implemented in the CASTEP package [23]. The generalized gradient approximation was used to account for exchange and correlation in the Perdew–Burke–Ernzerhof form [24], which is known to yield more accurate structural results [21], such as bond lengths, compared to the traditional local density approximation (LDA). The ionic positions have been optimized using the Broyden–Fletcher–Goldfarb–Shanno (BFGS) algorithm [25, 26]. The use of ultrasoft pseudopotentials [27] for all atoms has permitted us to exercise a lower plane-wave cutoff energy of 300 eV and hence achieve a shorter calculation time than would otherwise be possible in the case of norm-conserving pseudopotentials. A regular mesh of **k**-points, with typical dimensions of 5×5×3, was used to sample the irreducible Brillouin zone of the crystal. All the calculations correspond to $T = 0$ K. The convergence of the structural results with respect to the energy cutoff and the number of **k**-points was verified.

<u>The Landau theory and symmetry analysis.</u> Different structures observed in diffraction experiments are related via the corresponding irreducible representations (IRs) of the high-symmetry parent structure [28, 29, 30]. IRs define symmetry of the order parameter and the dimensionality of the order parameter space, together with a generic form of the Landau expansion of the free energy. Positions of the minima of the free energy determine the optimal values of the order parameters, resulting in a phase diagram that can be further mapped into the pressure-temperature coordinates.

**III. RESULTS AND DICUSSIONS**

**A. Structure of high pressure polymorph: study by diffraction and *ab-initio* modeling**

X-ray diffraction experiment as a function of pressure indicates that already at 1.1 GPa a new structural phase starts forming, and the transition completes at 1.4 GPa. The transition is

reversible – on decompression the new phase transforms back to the ambient conditions *I4mm* phase in the pressure interval 0.7-1.2 GPa. A dataset collected at 1.7 GPa has been used for symmetry assignment and structure solution. First 12 diffraction peaks were indexed by Dicvol [31] in a C-centered orthorhombic cell of nearly twice larger volume than of the ambient pressure *I4mm* structure.

Three space groups consistent with the observed systematic absences were considered: *Cmcm*, *C2cm*, *Cmc*$2_1$. The structure was first solved by global optimization in direct space using the program FOX [32]. A pronounced preferred orientation of crystallites with respect to the compression direction was detected in other light hydrides measured using identical diffraction geometry [33]. Therefore, one variable parameter modeling a possible preferred orientation (March-Dollase model) was also included into the optimization. The structure could be solved only in the space group *Cmc*$2_1$. The choice of the space group has been further supported by the Landau theory. The best fit was obtained for the direction of the preferred crystallite orientation <001>. All other directions yielded considerably worse fits.

Examination of the resulting structure and analysis by Platon [34] did not find any higher crystallographic symmetry. Non-hydrogen atoms were reliably located on the mirror plane, perpendicular to the *a* axis. Although the fit is somewhat sensitive to hydrogen atoms, their direct determination by global optimization does not appear to be reliable. However, a number of possible configurations of $NH_3$ and $BH_3$ groups is limited for *Cmc*$2_1$, so that each of the possible combinations can be examined in detail. Only 4 possible combinations of the orientations for $NH_3$ and $BH_3$ groups have been identified, considering their location on the mirror plane, and assuming that the H-atom substructure is ordered. Rotation of $NH_3$ or $BH_3$ by 180° around the B-N bond changes the eclipsed conformation of the $NH_3BH_3$ molecule into the staggered one. There is no uncertainty with assignment of N and B atoms: when they are swapped, the fit is affected significantly. However, 8 possible model structures were generated in order to verify all possible permutations. Each given model can be converted into any other one by applying the following transformations (or their combination):

- rotate $NH_3$ group around the B-N bond by 180°;
- rotate $BH_3$ group around the B-N bond by 180°;
- swap N and B atoms.

Four of these models have the eclipsed $NH_3BH_3$ conformation and the other four have the staggered one. These models have been further optimized using density functional theory (DFT) calculations, while keeping the experimental cell parameters fixed and then refined by Rietveld method against powder data using Fullprof [35]. The DFT geometry optimization yields 4 pairs of equivalent structures (numbers are assigned to the unique ones), which are ranked by energy in the Table 1. It is highly satisfactorily that the lowest energy structure gives also the best fit to the diffraction data.

TABLE 1. The summary for the DFT-optimized and Rietveld-refined $NH_3BH_3$ model structures. The structures were refined in the space group $Cmc2_1$ against the synchrotron diffraction data collected at 1.7 GPa.

| No. | Permutation | | | Conformation | Optimized energy per molecule,[a] eV | Bragg R-factor (%) | |
|---|---|---|---|---|---|---|---|
| | $NH_3$ rot | $BH_3$ rot | B-N swap | | | DFT-optimized structure | Rietveld-refined structure |
| | + | + | | Eclipsed | 0.289 | 49.3 | 14.5 |
| 4 | | + | | Eclipsed | 0.286 | 51.0 | 14.7 |
| | | + | | Staggered | 0.222 | 55.1 | 15.2 |
| 3 | | + | + | Staggered | 0.206 | 41.0 | 16.7 |
| 2 | + | + | + | Eclipsed | 0.104 | 13.2 | 8.5 |
| | | | | Eclipsed | 0.102 | 12.8 | 8.6 |
| | + | | | Staggered | 0.002 | 10.9 | 7.0 |
| 1 | + | | + | Staggered | 0 | 10.7 | 6.9 |

[a] Set to zero for the lowest energy structure.

The final refinement of the best structural model has been done as follows. The $z$ coordinate for the $BH_3$ group was fixed, thus defining the origin of the polar structure. The peak broadening was modeled by an orthorhombic strain, refining 3 parameters. The background was described by linear interpolation between selected points. The preferred orientation was modeled by one parameter (March's function), which converged to the value 0.772(9). In total, 40 reflections were fitted with only 4 intensity-dependent refined parameters. The refinement converged at $R_B$ = 6.9%, $R_F$ = 7.5%, $R_p$ = 12.5%, and $R_{wp}$ = 6.4%. The Rietveld refinement profile and the crystal structure are shown in Fig. 1. Atomic and cell parameters are listed in Table 2.

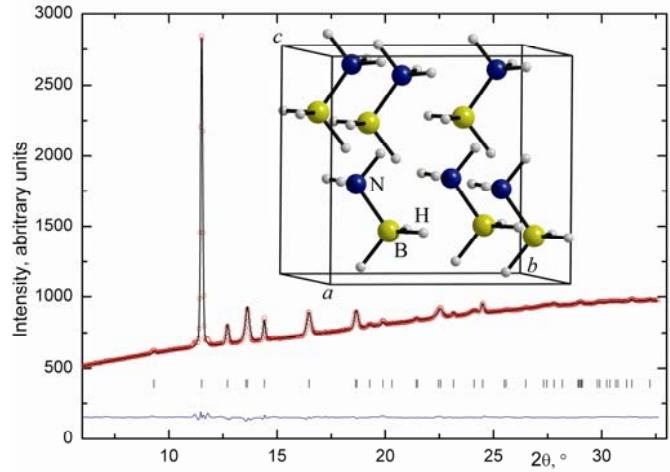

FIG. 1. (Color online) Rietveld refinement profile for the $Cmc2_1$ phase of $NH_3BH_3$ at 1.7 GPa, $\lambda$ = 0.71171 Å. The structure is shown in the inset.

TABLE 2. Experimental and DFT-optimized [in brackets] structural parameters for the high pressure polymorph of $NH_3BH_3$ at 1.7 GPa. DFT-optimized geometries of the $NH_3$ and $BH_3$ groups were retained during Rietveld refinement. The $z$ coordinate for the $BH_3$ group was fixed, thus defining the origin of the polar structure.

| Structure | Atom | x | y | z | B, Å² |
|---|---|---|---|---|---|
| $Cmc2_1$, Z = 4 | N1 | 0 | 0.2911(6) | 0.9176(13) | 4.9(2) |
| a = 5.9958(5) | | | [0.28158] | [0.93084] | |
| b = 6.4301(5) | B1 | 0 | 0.1529(10) | 0.70987 | 4.9(2) |
| c = 6.0293(7) Å | | | [0.15381] | [0.70987] | |
| | H1n | 0 | 0.1908 | 1.0516 | 5.9(2) |
| | | | [0.18122] | [1.06477] | |
| | H2n | 0.13913 | 0.3839 | 0.9350 | 5.9(2) |
| | | [0.13913] | [0.37430] | [0.94820] | |
| | H1b | 0 | 0.2689 | 0.55186 | 5.9(2) |
| | | | [0.26982] | [0.55186] | |
| | H2b | 0.16545 | 0.0439 | 0.70968 | 5.9(2) |
| | | [0.16545] | [0.04486] | [0.70968] | |

## B. Temperature and pressure evolution of the unit cell dimensions

Equipped with the knowledge of the structure of the discovered high-pressure $Cmc2_1$ phase, as well as two previously known ambient-pressure polymorphs of $NH_3BH_3$ [9⊥, 10⊥, 11⊥], we investigated their structural evolution as a function of the external stimuli. Here we show the temperature evolution of the ambient-pressure phases studied by *in-situ* synchrotron powder diffraction. The broad temperature range (110-300 K) and a fine temperature sampling (~1 K) were used to characterize in detail the highly anisotropic thermal expansion and to gain a further understanding of the mechanism of the phase transition. Cell dimensions obtained from the Rietveld refinement (the structural data from [10] were used as a starting model) of the 184 collected diffraction patterns, based on our image plate detector data, are shown in Fig. 2.

On cooling ammonia borane undergoes a structural phase transition at ~217 K, slightly lower than the previously reported transition temperature of 225 K. Our data confirm the *I4mm*-to-*Pmn*$2_1$ symmetry change and temperature dependences previously reported in [9, 13$^\perp$]. However, owing to the fine temperature sampling, a more detailed characterization of both phases becomes possible. The unit cell volume (Å$^3$) of the tetragonal phase depends linearly on temperature (K), $V = 130.79(3) + 0.02589(8)\,T$, suggesting a quasi-harmonic behaviour of the phonon system. The cell expansion is essentially isotropic for the tetragonal phase. The unit cell volume decreases at the *I4mm*-to-*Pmn*$2_1$ transition by 0.27% (Fig. 2). The two phases coexist in the narrow range 216-218 K. Below the transition temperature, variation of the cell parameters of the orthorhombic phase is highly anisotropic, and the temperature dependence of the unit cell volume is non-linear. There is no kink in the curve of the unit cell volume slightly above the transition temperature, suggested in Ref. [13]. Both the hysteresis and volume drop indicate the first order transition, which is however quite close to a second order, as follows from almost gradual evolution of the cell dimensions (Fig. 2). We did not observe an intermediate phase, suggested by Raman modes merging 10-12 K below the transition temperature [36]; the nearly second-order type of the transition is likely the origin of these observations. We did not observe either the hypothetical superstructure lines that may indicate partial ordering of the NH$_3$ and BH$_3$ groups in the tetragonal phase, suggested in Ref. [10]. The excellent statistics, exceeding in our experiment 10$^6$ counts per 2θ-step, would have allowed us to detect even very weak diffraction peaks.

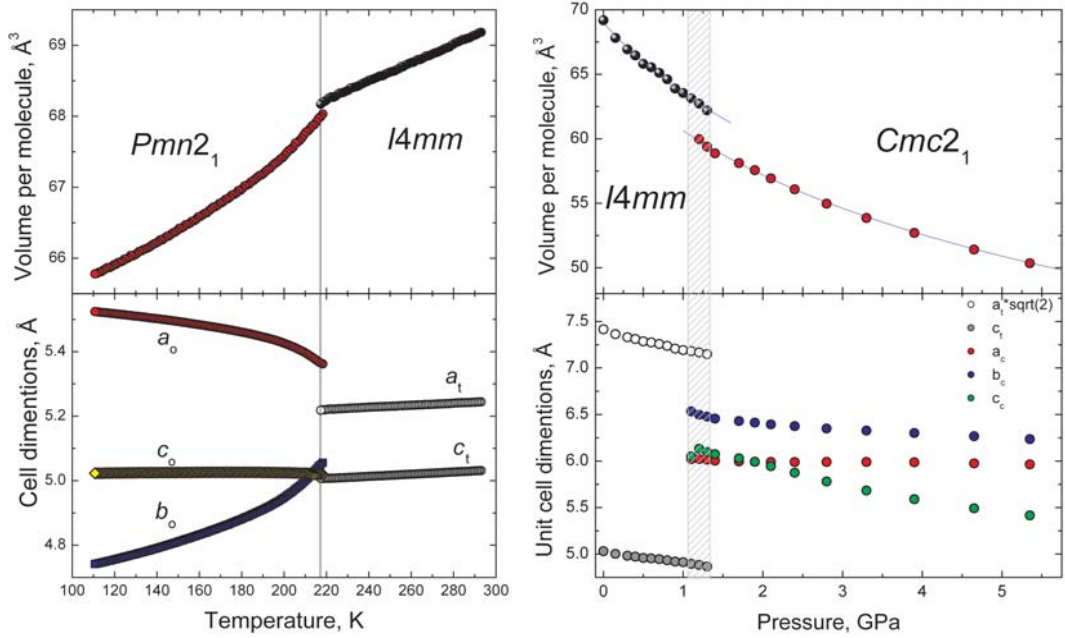

FIG. 2. (Color online) Variation of the molecular volume and unit cell dimensions as a function of temperature and pressure. The circles represent experimental data, and the lines are the best fits to the Murnaghan equation of state (right panel). Vertical line on the left panel indicates the transition temperature, there is a narrow (~2 K) temperature range where both phases coexist. The hatched region on the right panel corresponds to the apparent co-existence of the two phases as observed in the high-pressure experiment.

Under applied pressure at room temperature, the tetragonal phase transforms into another polymorph with $Cmc2_1$ symmetry. The pressure dependence of the Rietveld-refined unit cell parameters and the volume of $NH_3BH_3$ formula unit are shown in Fig. 2 [37]. The data were fitted with the Murnaghan equation of state, yielding the bulk moduli and their pressure derivatives:

$I4mm$: $B_0$ = 9.9(8) GPa, $B'_0$ = 4.8(14),

$Cmc2_1$: $B_0$ = 10.3(11) GPa, $B'_0$ = 4.6(4).

The new phase shows 4.4% volume contraction upon the transition. $V_0$ changes from 69.2(1) to 65.87(5) Å$^3$. While the cell contraction under pressure is nearly isotropic for the tetragonal phase, it is highly anisotropic for the high-pressure phase, with the $c$-axis parameter shrinking the most.

Our observation of a single pressure-induced transition, in the low-pressure range, is consistent with the Raman spectroscopic data in Ref. [14], but is inconsistent with Ref. [15], where two transitions were reported. The transition pressure of 1.1 GPa in our experiments with no pressure transmitting medium is slightly higher than 0.8 GPa in Ref. [14], where mineral oil was used. However it is expected that under non-hydrostatic conditions, the observed transition pressure is slightly higher than the true one, see Ref. [38] as an example. We did not detect any phase transition above 1.1 GPa and up to the maximum pressure of 12.1 GPa, as for instance the one at 5 GPa suggested from Raman spectroscopic data [16].

**C. Structural evolution under pressure and crystal chemistry of intermolecular interactions**

In the disordered tetragonal phase the $NH_3BH_3$ molecules are collinear with the $c$ axis. In the low-temperature $Pmn2_1$ phase on the other hand this arrangement becomes buckled – the molecules become slightly inclined to the $c$ axis. The angle between two molecules increases as the temperature is lowered: from 18° at 200 K slightly below the transition [11] to 26° at 90 K [10]. From our data we observe a smooth increase of the distortion in the $Pmn2_1$ phase as the temperature is lowered. In the high-pressure $Cmc2_1$ phase the inclination of $NH_3BH_3$ molecules with respect to the $c$ axis is much larger: the angle between two molecules jumps from zero to 69° upon the transition at 1.2 GPa and then increases linearly with pressure, up to 79° at 4.65 GPa. In this way at higher pressures the $NH_3BH_3$ molecules are situated almost perpendicularly to the $c$ axis (Fig. 1, inset). A very pronounced pressure-induced contraction along the $c$ axis is observed (Fig. 2) related to the increase of this angle.

The geometry of the strongly covalent $NH_3BH_3$ molecule is nearly invariant in the investigated pressure range. At 1.7 GPa it is practically the same as in the ambient-pressure phases: the DFT-optimized N-H and B-H distances are regular and equal to 1.03 Å and 1.21-1.215 Å, respectively. B-N-H angles are slightly larger (109.7-113.0°) than the N-B-H ones (107.7-110.3°), but both are close to tetrahedral coordination angle. The B-N bond is also robust under pressure: a small blue shift of the B-N stretching mode in the high-pressure Raman spectra suggests only a small shortening of the B-N bond on compression [14]. This is confirmed by our diffraction experiments: B-N bond in the $Cmc2_1$ phase shortens by only ~0.8% as the pressure is increased from 1.7 GPa to 4.65 GPa. Thus, the drastic pressure-driven changes involve mainly a reorganization of the intermolecular interactions.

Looking first at the heavier N and B atoms in the $Cmc2_1$ phase, one can notice that nitrogen atoms form a closed-packed structure, with all N…N distances exceeding 4 Å, while B…B distances show a wider spread and the corresponding network is more corrugated. The DFT-optimized structures with higher than optimal energy (Table 1) show shorter nearest-neighbour N…N distances. This suggests that the repulsive intermolecular $NH_3…H_3N$ interactions are stronger than $BH_3…H_3B$ ones. The bonding intermolecular $NH_3…H_3B$ interactions are represented by an extended pattern of dihydrogen N-H$^{\delta+}$…$^{\delta-}$H-B bonds, with their main geometric characteristics listed in Table 2. The H…H distances are considerably shorter than twice the value of the van der Waals radius of a hydrogen atom (2.4 Å). The N-

H…H groups of NH₃BH₃ tend to be linear while B-H...H tend to be bent, in a good agreement with an empirical rule established for a set of N-H…H-B containing structures [11]. We note, however, that the N-H vectors are not directly pointing to B or H atoms but rather to the B-H bonds. The dihydrogen bonds in NH$_3$BH$_3$ are longer and less directional than in the oxygen-containing systems, such as NaBH$_4$·2H$_2$O, where H…H distances at ambient pressure are as short as 1.77-1.79 Å [18, 39].

TABLE 3. Geometric characteristics of dihydrogen bonds in the $Cmc2_1$ high-pressure structure of NH$_3$BH$_3$.

| Distance, Å | 1.7 GPa | 4.65 GPa | Angle, ° | 1.7 GPa | 4.65 GPa |
|---|---|---|---|---|---|
| H1n…H2b | 1.97 | 2.01 | N1-H1n…H2b / B1 | 144 / 151 | 135 / 137 |
|  |  |  | B1-H2b…H1n | 91 | 92 |
| H2n…H2b | 2.15 | 1.81 | N1-H2n…H2b / B1 | 132 / 147 | 145 / 159 |
|  |  |  | B1-H2b…H2n | 137 | 144 |
| H2n…H2b | 2.03 | 2.05 | N1-H2n…H2b / B1 | 134 / 132 | 123 / 125 |
|  |  |  | B1-H2b…H2n | 109 | 109 |

The number of dihydrogen bonds in the low-temperature and the high-pressure phases is the same (12 H…H contacts per NH$_3$BH$_3$ molecule), and their length and directionality are similar. However the networks of the dihydrogen bond are quite different. While in the $Pmn2_1$ structure each H-atom takes part in two dihydrogen bonds (see Fig. 3a), in the HP phase one B-H bond is instead not involved in any H…H interactions and each of the other two B-H bonds is forming three H…H contacts (Fig. 3b).

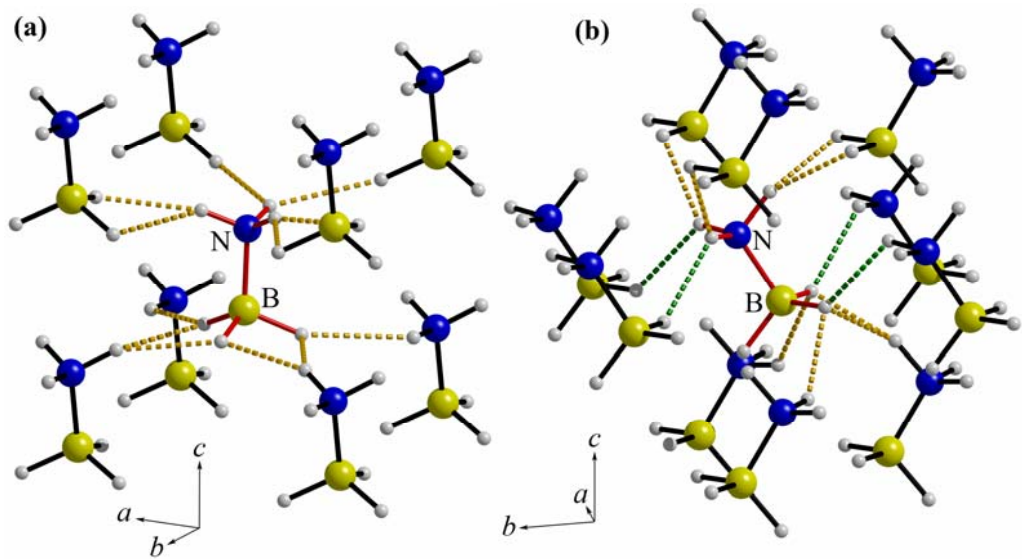

FIG. 3. Patterns of dihydrogen bonds in $Pmn2_1$ (a) and $Cmc2_1$ (b) phases of NH$_3$BH$_3$. Green (dark) dashed lines show the most compressible bonds, which combine molecules into layers situated in the $ab$ plane.

The difference between the two ordered phases appears to be much more pronounced when considering the structural architecture defined by the dihydrogen bonds. In the $Pmn2_1$ phase, each $NH_3BH_3$ molecule is associated with 8 other molecules enclosing the former in a cube (Fig. 3a). The resulting network of dihydrogen bonds is highly isotropic and represents a "frozen" pattern of H…H interactions existing in the disordered $I4mm$ phase. On the transition to the high-pressure phase we observe a dramatic change: only six of these eight molecules remain associated by dihydrogen bonds with the central unit. At the same time the large tilt of the collinear $NH_3BH_3$ molecules brings their ends closer and leads to a formation of new H…H contacts (shown as green dashed lines in Fig. 3b) involving 4 neighbouring molecules situated in the $ab$ plane. Thus, the more pronounced distortion in the high-pressure phase creates a different and more complex pattern of H…H interactions. As the pressure and the tilts of the $NH_3BH_3$ molecules are increased, the pattern of H…H bonds in the $Cmc2_1$ structure becomes more and more anisotropic: two types of dihydrogen bonds elongate and partly loose their directionality, while the dihydrogen bonds situated in the $ab$ plane become stronger and more linear (see Table 3) – at 4.65 GPa they strengthen to 1.81 Å. Markedly different compressibility of H…H bonds explains different pressure coefficients $\Delta\nu/\Delta P$ of the internal N-H and B-H vibrational modes, deduced from the high-pressure Raman studies of $NH_3BH_3$ [14, 15, 16]. Dihydrogen-bonded layers in the $ab$ plane manifest themselves by the anisotropic shift of peaks at pressures above 5 GPa [37] and they are possibly causing the observed preferred orientation in the [001] direction. The anisotropic compression of the unit cell (Fig. 2) and of the dihydrogen bonds is observed up to the maximum pressure of 12 GPa and it is expected to either saturate at higher pressures or result in another phase transition.

The energy difference between eclipsed and staggered conformations of $NH_3BH_3$ obtained by DFT calculations in the $Cmc2_1$ phase and shown in Table 1 are very close (0.1 eV per molecule = 9.6 kJ/mol) to those found for the free molecule, ~10.5 kJ/mol in Ref. [40] and ~8.4 kJ/mol in Ref. [41]. Thus the conformational differences are of the same energy scale as the dihydrogen bonding [18, 19]. The energy differences presented in the Table 1 are also relevant for the analysis of rotation dynamics. Table 4 summarizes geometrical details for the externally correlated 60° rotations of the $NH_3$ and $BH_3$ groups in the high-pressure phase. The energy differences between the models may be considered as activation energies of the corresponding rotations. Comparison of Tables 1 and 4 shows that three components contribute to the energy differences: short repulsive $NH_3…H_3N$ interactions (the most pronounced for the model 4), very short repulsive dihydrogen bonds (critical for the model 3), and more stable staggered versus eclipsed conformation. The highest energy barrier is observed for $NH_3$ rotation, the second highest is for the correlated rotation of the whole molecule, and the lowest one – for the rotation of the BH$_3$ group. This contrasts with the findings made for the low-temperature phase, where the barrier of the externally uncorrelated BH$_3$ rotation was found to be the highest, namely 38.3 kJ/mol from theory [17] and 23.6 kJ/mol from quasielastic neuron scattering [42].

TABLE 4. Geometry of intermolecular interaction in the rotational isomers of the $Cmc2_1$ structure at 1.7 GPa, determined by the DFT calculations. For the structural model numbers refer to the Table 1.

| No. | Shortest N…N distance, Å | Shortest N-H…H-B distance, Å | N-H…H angle, ° | B-H…H angle, ° | Calculated Pressure, GPa |
|---|---|---|---|---|---|
| 4 | 3.325 | 2.07 | 102 | 149 | 3.14 |
| 3 | 3.476 | 1.47 | 160 | 136 | 5.53 |
| 2 | 3.960 | 1.97 | 136 | 99 | 2.29 |
| 1 | 4.120 | 1.97 | 144 | 91 | 2.29 |

**D. Symmetry: diffraction and Landau theory. Phase diagram**

*1. Order parameter: symmetry and atomistic mechanism*

Having all the phases appearing as a function of temperature and pressure enumerated and crystal structures identified and solved, we apply a symmetry-based phenomenological analysis in order to understand atomistic mechanisms and thermodynamic aspects of the observed phase transitions, similar to the earlier study of another light hydride, LiBH$_4$ [43]. The embedding scheme for the unit cells of the three known structures of ammonia borane (Fig. 4) shows that both orthorhombic structures, low-temperature $Pmn2_1$ (basis vectors $a_0$, $b_0$, and $c_0$) and high-pressure $Cmc2_1$ ($a_p$, $b_p$, $c_c$) phases, are superstructures to the tetragonal $I4mm$ ($\mathbf{a_1}$, $\mathbf{a_2}$, $\mathbf{a_3}$). Their *primitive* (*P*) unit cell vectors are linked as following (induced strains are not included):

$$a_0 = \mathbf{a_2} + \mathbf{a_3}, \quad b_0 = \mathbf{a_1} + \mathbf{a_3}, \quad c_0 = \mathbf{a_1} + \mathbf{a_2};$$
$$a_p = \mathbf{a_2} + \mathbf{a_3}, \quad b_p = \mathbf{a_1} + \mathbf{a_3}, \quad c_p = \mathbf{a_1} + \mathbf{a_2},$$
$$(a_c = a_p + b_p, \quad b_c = -a_p + b_p). \tag{1}$$

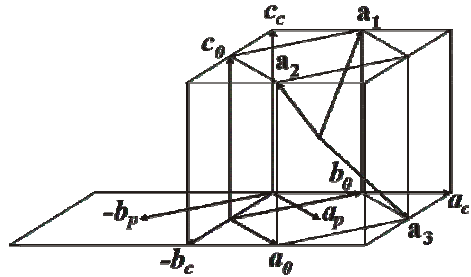

FIG. 4. Relation between primitive unit cells of $I4mm$ ($\mathbf{a_i}$), low-temperature $Pmn2_1$ ($a_0$, $b_0$, $c_0$), and high-pressure $Cmc2_1$ ($a_p$, $b_p$, $c_c$) structures. The C-centered orthorhombic Bravais cell ($a_c$, $b_c$, $c_c$) is also shown.

The symmetry breaking of the lattice translations [Eqs. (1)] is described, therefore, by the same vector star $\mathbf{k}=(\mathbf{a}^*_1+\mathbf{a}^*_2+\mathbf{a}^*_3)/2$ of the corresponding reciprocal space. All vectors $\mathbf{a}_i$, $\mathbf{b}_i$, and $\mathbf{c}_i$ from (1) satisfy the condition $\exp(\mathbf{k}\mathbf{a}_i)=1$ which is necessary and sufficient for vectors $\mathbf{a}_i$ to belong to the same translational subgroup of the parent tetragonal lattice.

Straightforward group-theoretical procedure (see, for example, Ref. [30]) identifies a single two-component order parameter (OP), which combines ordering and atomic displacement mechanisms, for both the transformations $I4mm(Z=1)$–$Pmn2_1(Z=2)$ and $I4mm(Z=1)$–$Cmc2_1(Z=2)$, where Z is the number of formula units in the *primitive* cell. The OP transforms as the two-dimensional irreducible representation $M_5$ (in notations of Ref. [30]), which is isomorphic to the vector representation of the point group 4*mm*. The corresponding non-equilibrium thermodynamic potential has four minima: one corresponds to the orientationally disordered parent phase *I4mm*, and three − to the ordered distorted structures: $Cmc2_1$, $Pmn2_1$, $P2_1$ [30]. Three out of the four mentioned phases: orientationally disordered tetragonal and two ordered orthorhombic phases were indeed observed in $NH_3BH_3$, whereas possible existence of the monoclinic $P2_1$ phase follows from our analysis.

Figure 5 shows displacement components of the structure distortions transforming the parent tetragonal phase of $NH_3BH_3$ to (i) HP orthorhombic $Cmc2_1$ – a single OP component is unequal to zero ($\eta_1\neq 0$, $\eta_2=0$), (ii) LT $Pmn2_1$ – two non-zero OP components are equal ($\eta_1=\eta_2\neq 0$). A monoclinic $P2_1$ structure would correspond to two unequal OP components ($\eta_1\neq 0$, $\eta_2\neq 0$). Both observed transformations induce onset, as secondary order parameters, of spontaneous strains: $e_{11} = 0.037$, $e_{22} = -0.054$, $e_{33} = 0.005$ at the *I4mm* to $Pmn2_1$ transition, and $e_{11} = e_{22} = -0.159$, $e_{33} = 0.20$, $e_{12} = 0.04$ at *I4mm* to $Cmc2_1$ one; the values are calculated from the experimental unit cell dimensions. Therefore, this compound belongs to the class of improper ferroelastics.

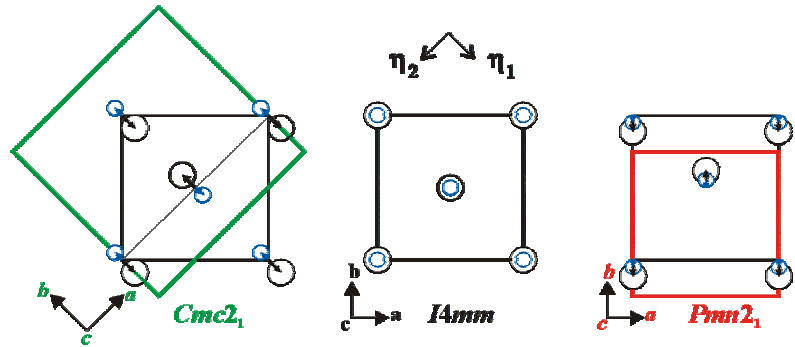

FIG. 5. (Color online) Atomic displacements in the orthorhombic structures of $NH_3BH_3$. Coordinate system ($\eta_1$, $\eta_2$) for the components of the order parameter is shown for reference.

## 2. Landau free energy and phase diagram

The thermodynamic potential corresponding to the two-component order parameter ($\eta_1,\eta_2$) of the aforementioned irreducible representation $M_5$ can be written generally as a Landau free energy expansion in powers of $\eta_1$, $\eta_2$:

$$F_0(\eta_1,\eta_2) = \alpha_1(\eta_1^2+\eta_2^2)+\alpha_2(\eta_1^2+\eta_2^2)^2+\beta_1(\eta_1^4+\eta_2^4)+\beta_2(\eta_1^4+\eta_2^4)^2+\kappa(\eta_1^2+\eta_2^2)(\eta_1^4+\eta_2^4). \quad (2)$$

An 8$^{th}$-order term has to be included in order to ensure that all solutions of the equations of state $\partial F_0/\partial \eta_i=0$ correspond to energy minima (including the observed HP and LT phases and the predicted monoclinic $P2_1$ phase). From two possible 6$^{th}$-order terms we keep only one, a product of the basis invariants, which tunes the plane section of the theoretical phase diagram to experimental observations. One finds a generic phenomenological phase diagram for the Landau potential (2) in Refs. [28, 29]. It is worth noting that the potential (2) is a particular case for more general model (see Refs. [28,29]) corresponding to the OP which belongs to the image group B8a in notations of Ref. [30].

Figure 6 maps experimental data of the pressure-temperature induced transformations in NH$_3$BH$_3$ onto theoretical phase diagram calculated in the phase space of the coefficients of the Landau free energy (2). The key role is played by the coefficients $\alpha_1$ and $\alpha_2$ of the above expansion, which are plotted on the axes in Fig. 6, and which correspond, in the rotated coordinate system, to the pressure and temperature axes. It turns out that two different generic phase diagrams can be obtained from the Landau potential (2), depending on the ratio of the coefficients ($4\alpha_2\beta_2/\kappa^2$), and are depicted in Figs. 6a) and 6b) respectively.

Despite the entirely phenomenological nature of the expansion (2), certain conclusions can be drawn by relating theoretical phase diagram in Fig. 6 to the experimental data. First, we can expect the existence of tricritical points on the phase diagram and, therefore, a line of continuous phase transitions between the parent tetragonal phase and the low-symmetry orthorhombic ones. Both experimentally observed transformations, temperature- and pressure-driven, are discontinuous, which makes us put the corresponding points $T_C$ and $P_C$ beyond the tricritical points $Tr$ predicted by theory (Fig. 6). However, by varying external parameters, the thermodynamic pathway can be tuned to a continuous regime for these transformations. The latter prediction is supported by the results of isotopic H/D substitution in the NH$_3$BH$_3$ structure, where it was found that in the fully deuterated ND$_3$BD$_3$ sample, the temperature-induced *I4mm*-to-*Pmn*2$_1$ transition occurs very close to the second-order regime [13].

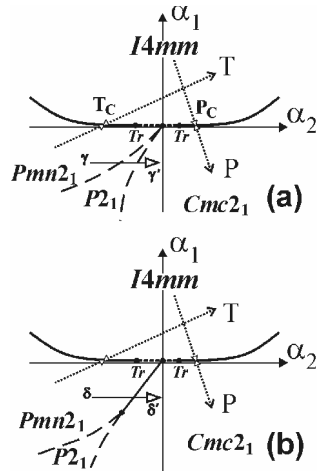

FIG. 6. Equilibrium phase diagrams for the Landau potential (2): (a) $\alpha_2 > (\kappa^2/4\beta_2)$; (b) $0 < \alpha_2 < (\kappa^2/4\beta_2)$. Dashed and solid lines correspond to second- and first-order phase transitions. $Tr$ are tricritical points. Temperature-pressure axes are schematically shown by dotted lines.

The second prediction of the Landau theory concerns the possible existence of an intermediate monoclinic $P2_1$ phase, located on the P-T diagram between the $Pmn2_1$ and $Cmc2_1$ phases. Indeed, theoretical diagrams indicate at least two possible topologies for the phase transition paths: (a) transformation of the $Pmn2_1$ structure, under non-ambient conditions, to the $Cmc2_1$ phase can be realized by two continuous transitions through the hypothetical $P2_1$ phase [$\gamma$-$\gamma'$ path on Fig. 6(a)], or (b) a direct discontinuous transformation distorts one orthorhombic structure into the other [Fig. 6(b), $\delta$-$\delta'$ path]. Note that the orthorhombic phases of $NH_3BH_3$ both have ordered structures, in full agreement with the results of the group-theoretical analysis. It means, in turn, that possible phase transitions between these phases must be of the displacement type.

### 3. Symmetry induced coupling

As a phenomenological construction, the Landau potential (2) does not reveal the microscopic origin of the coefficients $\alpha_i$, $\beta_i$, $\kappa$ ($i=1,2$). The purpose of this section is to elucidate their origin by appealing to the structural and symmetry arguments, and in particular write down explicitly the coupling between the order parameters and the elastic properties of the crystal (strain tensor). Indeed, the theoretical phase diagrams in Fig. 6 are entirely determined by the symmetry of the primary transition mechanisms, namely the orientational ordering of $NH_3$ and $BH_3$ pyramids and their simultaneous displacements (see Fig. 5). To describe anomalies in the behavior of macroscopic deformations observed in our XRD experiments, the model of Eq. (2) should be complemented with a coupling between the primary order parameter and the elastic subsystem.

On the theoretical side, one more important modification should be made since the mapping of the experimental data onto the theoretical phase diagrams (Fig. 6) experiences a certain difficulty, typical of a classical phenomenological theory. Namely, one may be tempted to conclude, from analyzing Fig. 6, that at least two phenomenological coefficients, $\alpha_1$ and $\alpha_2$, must be temperature and pressure dependent. This is in contrast to the conventional hypothesis, which establishes all the coefficients $\alpha_n$ of the anharmonic terms ($\alpha_n \eta^n$ $n > 2$) in the Landau potential to be constant, and only the quadratic coefficient $\alpha_1$ to be a linear function of external thermodynamic parameters. Nevertheless, this latter complexity does not affect the coupling between primary and secondary order parameters, and this is of great importance from the experimental point of view. Indeed, it is an experimental challenge to find directly, using X-ray diffraction on $NH_3BH_3$, the displacements of light atoms, which constitute a primary order parameter, especially at non-ambient conditions with a sample placed in a diamond anvil cell. By contrast, it is a much easier task to follow the variation of the lattice parameters with pressure, since elastic strains, calculated from the lattice parameters, provide the temperature/pressure dependences of the corresponding OPs via the symmetry allowed coupling.

In the light of the above discussion, let us consider a modified Landau potential with explicit dependence on the components of the stress tensor $e_i$, which play the role of the secondary order parameters:

$$F(\eta, e_i) = F_0(\eta) + F_1(e_i) + F_2(\eta, e_i), \qquad (3)$$

which contains a non-critical contribution $F_1(e_i)$ and the coupling energy $F_2(\eta, e_i)$. We use hereafter the Voigt notations for the strain components: $e_1 = e_{11}$, $e_6 = e_{12}$. For $F_1(e_i)$ we consider orthorhombic strains $e_1 = -e_2$ and $e_6$, and restrict the expansion to the second order terms:

$$F_1(e_1, e_6) = \tfrac{1}{2} c_{11} e_1^2 + \tfrac{1}{2} c_{66} e_6^2, \qquad (4)$$

where $c_{11}$ and $c_{66}$ are tetragonal stiffness constants. The lowest order coupling that satisfies the symmetry conditions is

$$F_2(\eta, e_1, e_6) = \gamma_1 \eta_1 \eta_2 e_1 + \gamma_2 e_1^2 (\eta_1^2 + \eta_2^2) + \gamma_3 (\eta_1^2 - \eta_2^2) e_6. \qquad (5)$$

The biquadratic coupling term $\gamma_2$ is also included in $F_2$ due to the onset of non-negligible orthorhombic distortions in the high pressure phase $Cmc2_1$ (see Sec. III.C.1). The minimization of the potential $F$ with respect to $\eta_i$ and $e_i$ yields, for equilibrium orthorhombic deformation in the low temperature (LT) and high-pressure (HP) phases:

$$\text{LT } (\eta_1=\eta_2=\eta\neq 0): \quad e_{1(T)} = -\frac{\gamma_1}{c_{11}}\eta_1\eta_2 = -\frac{\gamma_1}{c_{11}}\eta^2, \quad e_{6(T)} = -\frac{\gamma_3}{c_{66}}\left(\eta_1^2 - \eta_2^2\right) = 0;$$

$$\text{HP } (\eta_1\neq 0, \eta_2=0): \quad e_{1(P)}^2 = -\frac{1}{\gamma_2}\left[\alpha_1 - (2\alpha_2 + \beta_1)\eta^2\right], \quad e_{6(P)} = -\frac{\gamma_3}{c_{66}}\eta^2. \quad (6)$$

Several conclusions from the equations (6) can be made to compare with the available experimental data. First, the prediction of the absence of a share distortion $e_{6(T)}$ in the LT phase with $Pmn2_1$ symmetry perfectly corresponds to our data. Secondly, one finds for the LT phase an induced strain $e_{1(T)}$ to be proportional to $\eta^2$, which is characteristic for an *improper ferroelastic* transformation. Finally, equations for the strains $e_{1(P)}$ and $e_{6(P)}$ induced in the high pressure structure yield the relation $e_{1(P)}^2 \sim e_{6(P)}$. Thus, irrespective of the complexity of the temperature and pressure dependences of the phenomenological coefficients $\alpha_i$ ($\beta_i$) in the corresponding Landau potential, the above functional relations between the primary and secondary OPs must be satisfied.

The microscopic character of the displacement component of the order parameters $\eta_i$ is rather simple as seen from Fig. 5. Here we use the evolution of the *y*-coordinate of B atom as a measure of the corresponding OP. Figure 7 illustrates some of the observed couplings exemplified in plots of $e_{1(T)}$ vs. $\eta^2$, and $e_{1(P)}^2$ vs. $e_{6(P)}$. Their linear character perfectly corresponds to the above predictions of the phenomenological theory with the model potential given in Eq.(3-5). Having this model potential and assuming hydrogen atoms to be rigidly bound to nitrogen and boron, the shifts of all atoms in $NH_3BH_3$ as a function of temperature and pressure can be well described merely on the basis of the unit cell deformations.

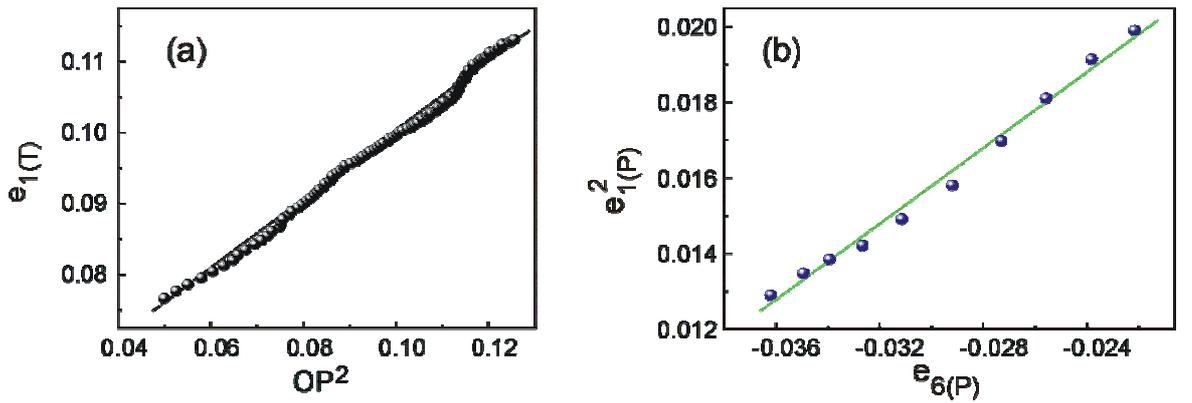

FIG. 7. (Color online) (a) Coupling between the structural order parameter (taken as a *y*-coordinate of B atom) and the spontaneous elastic strain $e_{1(T)}$ in the LT phase $Pmn2_1$, in very good agreement with the prediction of the Landau theory (dashed line); (b) Squared spontaneous strain $e_{1(P)}$ vs. $e_{6(P)}$ in HP phase $Cmc2_1$. Straight lines are the best least square fits based on Eq.(6).

## IV. CONCLUSIONS

In summary, we have characterized the phase transition phenomena in ammonia borane by a combination of experimental and theoretical techniques. We have found a new, stable under pressure polymorph and characterized its crystal structure. The symmetry and positions of the B and N atoms were found from diffraction data. A set of possible structures with respect to the location of H atoms was generated. This model structures were optimized with the help of first-principles DFT calculations and then refined by the Rietveld method, unambiguously identifying the correct solution. Such a combination of methods reveals not only the symmetry of the new phase, but also provides positions of the hydrogen atoms which are notoriously difficult to extract from the high-pressure X-ray powder diffraction data. On the other hand, theoretical structure prediction when used alone often does not provide correct crystal structure and properties of light hydrides (for the recent review see Ref. [44] and for the more general discussion Ref. [45]).

A set of basic properties of ammonia borane, such as compressibility and thermal expansion, have also been measured. Group-theoretical analysis identifies a single two-component order parameter, combining ordering and atomic displacement mechanisms, which link an orientationally disordered parent phase *I4mm* with ordered distorted *Cmc*$2_1$, *Pmn*$2_1$ and *P*$2_1$ structures. We propose a generic phase diagram for $NH_3BH_3$, mapping three experimentally found and one predicted (*P*$2_1$) phases as a function of temperature and pressure, along with the evolution of the corresponding structural distortions. We have shown that ammonia borane belongs to the class of improper ferroelastics; both temperature- and pressure-induced phase transitions are allowed to be of the second order.

Our data illustrate an important role of N-H…H-B dihydrogen bonds and other intermolecular interactions in the stability of $NH_3BH_3$ polymorphs. In the ambient-pressure phases, the patterns of dihydrogen bonds are very similar. In the high-pressure phase on the other hand, large tilts of $NH_3BH_3$ molecules with respect to the *c* axis result in a different and more complex pattern of H…H interactions. As the pressure is increased, two types of dihydrogen bonds elongate and partly loose their directionality, while the dihydrogen bonds situated in the *ab* plane become stronger and more linear, thus creating dihydrogen-bonded layers. Our DFT calculations quantify the influence of intermolecular interactions on the $NH_3$ and $BH_3$ rotational energy barriers in the high-pressure phase and show that the highest energy barrier is observed for the $NH_3$ rotation, the second highest – for the correlated rotation of the molecule as a whole, and the lowest one – for the rotation of the $BH_3$ group.

Finally, we would like to comment on the power of the combined experimental and theoretical methodology used in the present work. Careful use of the first-principles computational methods together with global optimization in direct space and the Rietveld refinement has proven to be a very accurate and powerful method of characterizing novel structures, in particular those stabilized under high pressure. In addition, symmetry-based thermodynamic analysis combined with the powder diffraction experiment allows to construct a pressure-temperature phase diagram and even go beyond to make predictions of new possible phases. In particular, symmetry-restricted couplings between the structural distortions and spontaneous strains give access to the atomic shifts related to the order parameter. These strains are explicitly related to the temperature and pressure evolution of the unit cell dimensions. Such a combination of theory and experiment is very helpful to the high-pressure diffraction experiments, where the quality of the data may prove insufficient to allow precise calculation of atomic coordinates, whereas the unit cell dimensions as a function of pressure can be readily extracted. The basic characteristics, such as the critical temperature and pressure of the phase transitions, as well as compressibility and thermal expansion in different phases still have to be obtained from the experiment. Sensible calculation of these properties could be considered as a future goal for various *ab-initio* models, and should hopefully assist in realistic predictions of new polymorphs and their stability.

## V. ACKNOWLEDGEMENTS

We acknowledge SNBL for in-house beam time allocation.

# References

* Yaroslav.Filinchuk@esrf.fr